\documentclass[12pt]{elsart}
\usepackage{graphicx}
\usepackage{amssymb}
\usepackage{times}
\begin{document}

\newcommand{\re}{\mathop{\mathrm{Re}}}
\newcommand{\im}{\mathop{\mathrm{Im}}}
\newcommand{\D}{\mathop{\mathrm{d}}}
\newcommand{\I}{\mathop{\mathrm{i}}}

\def\lambar{\lambda \hspace*{-5pt}{\rule [5pt]{4pt}{0.3pt}} \hspace*{1pt}}

\noindent {\Large DESY 10-048}

\noindent {\Large March 2010}

\bigskip

\begin{frontmatter}

\journal{Phys. Rev. ST-AB}

\date{}

\title{Using the longitudinal space charge instability for generation of
VUV and X-ray radiation}

\author{E.A.~Schneidmiller}
\author{and M.V.~Yurkov}

\address{Deutsches Elektronen-Synchrotron (DESY),
Notkestrasse 85, D-22607 Hamburg, Germany}

\begin{abstract}

Longitudinal space charge (LSC) driven microbunching instability in electron beam formation
systems of X-ray FELs is a recently discovered effect hampering beam instrumentation and
FEL operation. The instability was observed in different facilities
in infrared and visible wavelength ranges.
In this paper we propose to use such an instability
for generation of VUV and X-ray radiation. A typical longitudinal space charge amplifier (LSCA)
consists of few amplification cascades
(drift space plus chicane) with a short undulator behind the last cascade. If the amplifier starts up
from the shot noise, the amplified density modulation has a wide band, on the order of unity.
The bandwidth of the radiation within the central cone is given by inverse number of undulator periods.
A wavelength compression could be an attractive option for LSCA since the process is broadband, and
a high compression stability is not required.
LSCA can be used as a cheap addition to the existing or planned short-wavelength FELs.
In particular, it can produce the second color for a pump-probe experiment. It is also
possible to generate attosecond pulses in the VUV and X-ray regimes.
Some user experiments can profit from a relatively large bandwidth of the radiation, and this is easy
to obtain in LSCA scheme.
Finally, since the amplification mechanism is broadband and robust, LSCA can be an interesting alternative to
self-amplified spontaneous emission free electron laser (SASE FEL) in the case of using laser-plasma
accelerators as drivers of light sources.
\end{abstract}

\end{frontmatter}

\bigskip

\bigskip

\noindent Prepared for the Workshop on the Microbunching Instability III,
Frascati, 24-26 March 2010

\baselineskip 20pt

\clearpage

\section{Introduction}

Longitudinal space charge (LSC) driven microbunching instability \cite{fel-bc,lsc-mb} in electron linacs with
bunch compressors (used as drivers of short wavelength FELs) was a subject of intense theoretical and experimental
studies during last years \cite{huang-2004,venturini,huang-chao,loos,schmidt,lumpkin,heater-oper}. Such
instability develops in infrared and visible wavelength ranges and can hamper electron beam diagnostics and FEL operation.

In this paper we propose to use this effect for generation of VUV and X-ray radiation. We introduce a concept of an
LSC amplifier and present basic scaling relations in Section 2. We discuss possible applications of
such an amplifier in Section 3, and end up with discussion and conclusions in Section 4.

\section{Generic LSC amplifier}

\begin{figure*}[b]

\includegraphics[width=1.0\textwidth]{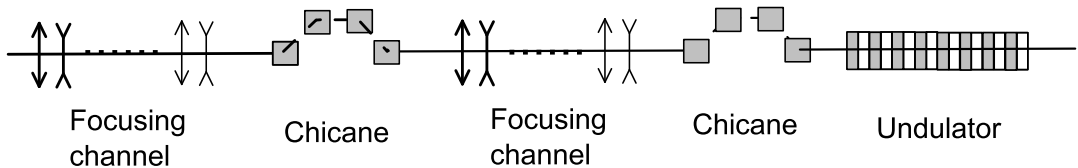}

\caption{\small Conceptual scheme of an LSC amplifier.}

\label{scheme}
\end{figure*}

\subsection{Scheme of an LSCA}

Let us consider a scheme presented at Fig.~\ref{scheme}. An amplification cascade consists of a focusing
channel and a dispersive element (usually a chicane) with an optimized momentum compaction $R_{56}$.
In a channel the energy modulations are accumulated, that are proportional to density modulations and
space charge impedance of the drift space. In the chicane these energy modulations are converted into
induced density modulations that are much larger that initial ones \cite{fel-bc}, the ratio defines a gain per
cascade. In this paper we will mainly consider the case when the amplification starts up from the shot noise in
the electron beam (although, in principle, the coherent density modulations can be amplified in the same way).
A number of cascades
is defined by the condition that the total gain (product of partial gains in each cascade) is sufficient
for saturation (density modulation on the order of unity) after the start up from shot noise. As we
will see, in most cases two or three cascades would be sufficient. The amplified density modulation has
a large relative bandwidth, typically in the range 50-100 \%.
Behind the last cascade a radiator undulator is installed, which
produces a powerful radiation with a relatively narrow line
(inverse number of periods) within the central cone. This radiation is transversely coherent,
and the longitudinal
coherence length is given by the product of the number of undulator periods by the radiation wavelength.
When LSCA saturates in the last cascade, a typical enhancement of the radiation intensity
over spontaneous emission is given by a number of electrons per wavelength.

\subsection{Formula for a gain per cascade}

Let us now present simple formulas for calculations of the gain and optimization of parameters of an
LSCA. As in the case of a SASE FEL \cite{ks-sase},
we assume that at the entrance to the amplifier there is only shot noise in the electron beam.
Let us consider the linear amplification of spectral components of the noise within the amplifier band.
The formula for amplitude gain per cascade
was obtained in studies of microbunching instabilities in linacs with
bunch compressors \cite{fel-bc}:

\begin{equation}
G_n = Ck|R_{56}|
\frac{I}{\gamma I_{\mathrm{A}}}
\frac{4\pi|Z(k)| L_d}{Z_0}
\exp \left(-\frac{1}{2}C^2k^2R_{56}^2\frac{\sigma_{\gamma}^2}{\gamma^2}
\right) \ .
\label{gain-wake}
\end{equation}

\noindent Here $k= 1/\lambar = 2\pi/\lambda$ is the modulation wavenumber,
$Z$ is the impedance of a drift space (per unit length),
$Z_0$ is the free-space impedance, $L_d$ is the length of the drift space,
$I$ is the beam current,
$I_{\mathrm{A}}$ is the Alfven current, $R_{56}$ is the compaction factor of a chicane,
$C$ is the compression factor\footnote{The wavenumber after the chicane is $Ck$.}, $\gamma$ is
relativistic factor, and $\sigma_{\gamma}$ is rms uncorrelated energy
spread (in units of rest energy). It is assumed here that energy modulations are accumulated upstream of
the chicane but not inside: there the self-interaction is suppressed due
to $R_{51}$ and $R_{52}$ effects \cite{we-micr,stup-micr,kim-micr}.
We also assume in the following that a length of the drift space is much larger than that of the
chicane, while the $R_{56}$ of the drift space is much smaller.
Also note that the formula (\ref{gain-wake}) was obtained under the condition of high gain, $G \gg 1$.
In this case both the sign of $R_{56}$ and the phase of impedance are not important. The value of the $R_{56}$
is to be optimized for a highest gain at a desired wavelength.

In this paper we will consider LSC induced energy modulations in a drift space
or in an undulator (at a
wavelength that is much longer than the resonant one). In the latter case\footnote{Under the condition
$\sigma_{\bot}^2 \ll \lambar \lambar_w$ ,where $\lambar_w = \lambda _w/2\pi $ is the reduced undulator period and
$\sigma_{\bot}$ is rms transverse size of the beam.}
the relativistic factor $\gamma$ in
the formulas for impedance should be substituted by the longitudinal relativistic factor $\gamma_z$
\cite{geloni-und-wake}:

\begin{equation}
\frac{4\pi Z(k)}{Z_0} =
\frac{2 i k}{\gamma_z^2} \int d \vec{r_{\bot}'} \int d \vec{r_{\bot}''}
\rho(\vec{r_{\bot}'}) \rho(\vec{r_{\bot}''})
K_0 \left( \frac{k |\vec{r_{\bot}'}-\vec{r_{\bot}''}|}{\gamma_z} \right) \ .
\label{imp-geloni}
\end{equation}

\noindent Here $\rho(\vec{r_{\bot}})$ is the transverse distribution of the current density
and $K_0$ is the modified Bessel function.

We requiest that a density modulation does not change significantly in the drift space.
This may happen due to plasma oscillations or due to a spread of longitudinal velocities for
a finite beam emittance. Thus, a drift space length is limited by the condition:

\begin{equation}
L_d \le min(L_1,L_2) \ .
\label{ld-max}
\end{equation}

\noindent Here $L_1$ is the reduced wavelength $\lambar_p$ of plasma oscillations:

\begin{equation}
L_1 \simeq \lambar_p = \gamma_z \left( \frac{I}{\gamma I_{\mathrm{A}}} \frac{4 \pi |Z| k}{Z_0} \right)^{-1/2}
\ .
\label{l1}
\end{equation}

\noindent The second limitation follows from the condition that the longitudinal velocity spread
due to a finite beam
emittance does not spoil the modulations during the passage of the drift:

\begin{equation}
L_2 \simeq \frac{\lambar}{\sigma_{\theta}^2} = \frac{\beta \lambar}{\epsilon} \ ,
\label{l2}
\end{equation}

\noindent where $\sigma_{\theta}$ is the angular spread in the beam, $\beta$ is the beta function,
$\epsilon = \epsilon_{n}/\gamma$ is beam emittance, $\epsilon_{n}$ is the normalized emittance.

\subsection{Formulas for an optimized LSC amplifier}

Let us first consider the case without wavelength compression, $C=1$. We start optimization assuming that
the beam parameters are fixed: current $I$, normalized emittance $\epsilon_{n}$,
beam energy $\gamma$ and energy spread $\sigma_{\gamma}$ in units of the rest energy, and longitudinal
gamma-factor $\gamma_z$. We can select a central wavelength,
optimize $R_{56}$ of the dispersion section for the chosen wavelength, choose
beta-function and optimize a length of the drift space. Our goal is to get a highest gain at a
shortest wavelength.

The impedance increases with $k$, achieves the
maximum at

\begin{equation}
\lambar \simeq \lambar_{opt} \simeq \frac{\sigma_{\bot}}{\gamma_z} = \frac{\sqrt{\epsilon \beta}}{\gamma_z} \ ,
\label{lambar-opt}
\end{equation}

\noindent and then decays in the asymptote of a pancake beam. Let us consider the wavelength
about $2\pi \lambar_{opt}$ as an optimum choice, since the impedance is the largest, and transverse correlations
of the LSC field are still on the order of the beam size. The impedance at this wavelength can be
approximated by

\begin{equation}
\frac{4\pi |Z|}{Z_0} \simeq \frac{1}{\lambar \gamma_{z}^2} \simeq
\frac{1}{\sigma_{\bot} \gamma_{z}} \ .
\label{impedance-opt}
\end{equation}

\noindent It is slightly underestimated, but we ignore here numerical factors on the order
of unity.
Note that the impedance is rather flat around the optimal wavelength, it
reduces by 10-15 \% when the wavelength is by a factor of 2 smaller or larger than the optimal one.
The optimal $R_{56}$ of the dispersion section is
\begin{equation}
R_{56} \simeq \lambar \frac{\gamma}{\sigma_{\gamma}} \ .
\label{r56-opt}
\end{equation}

\noindent Substituting (\ref{impedance-opt}) and (\ref{r56-opt}) into (\ref{gain-wake}), we get an estimate of
the amplitude gain per cascade for the wavelength given by (\ref{lambar-opt}):

\begin{equation}
G_n \simeq \frac{I}{\sigma_{\gamma} I_{\mathrm{A}}} \
\frac{L_d}{\lambar \gamma_z^2} \ .
\label{gain-ld}
\end{equation}

\noindent Thus, the gain per cascade is approximately equal to the
longitudinal brightness of the electron beam multiplied by a number of LSC formation lengths.

Let us now consider the limitations on the drift length (\ref{ld-max}). The first limit (\ref{l1}) can
be rewritten with the help of (\ref{impedance-opt}) as:

\begin{equation}
L_1 \simeq \lambar_p \simeq \lambar \gamma_z^2  \sqrt{\frac{\gamma I_{\mathrm{A}}}{I}} \ .
\label{l1-opt}
\end{equation}

\noindent So, in the case $L_1 < L_2$ the drift space length can be chosen to be $L_d \simeq L_1$. In
this case the expression for the gain (\ref{gain-ld}) reduces to:

\begin{equation}
G_n \simeq \frac{1}{\sigma_{\gamma}} \sqrt{\frac{\gamma I}{I_{\mathrm{A}}}} \ .
\label{gain-l1}
\end{equation}

\noindent This estimate for the gain was originally obtained in \cite{fel-bc}. In the considered limit we have a
relatively large beta-function. It is advisable to reduce it (if technically possible), since
the wavelength (\ref{lambar-opt}) and length of the drift
space (\ref{l1-opt}) are also reduced but the gain (\ref{gain-l1}) stays the same. This happens until the
spread of longitudinal velocities starts playing a role, i.e. when $L_1 \simeq L_2$. The corresponding
beta-function is

\begin{equation}
\beta_{cr} \simeq \epsilon \gamma_z^2  \sqrt{\frac{\gamma I_{\mathrm{A}}}{I}} \ .
\label{beta-opt}
\end{equation}

\noindent If one further reduces beta-function, $\beta < \beta_{cr}$,
the  maximal drift length is given by $L_d \simeq L_2$.
In this case from (\ref{l2}), (\ref{lambar-opt}), and (\ref{gain-ld}) we find that
the gain is proportional to the beam brightness in 6-D phase space:

\begin{equation}
G_n \simeq \frac{I}{\sigma_{\gamma} I_{\mathrm{A}}} \
\frac{\beta}{\epsilon \gamma_z^2} \simeq
\frac{I}{\sigma_{\gamma} I_{\mathrm{A}}} \
\left( \frac{\lambar}{\epsilon} \right)^2 \ .
\label{gain-l2}
\end{equation}

\noindent Although the gain can still be high for $\lambar \gg \epsilon$, it quickly decreases when
one goes to shorter wavelengths - contrary to the case (\ref{gain-l1}). Thus, the condition
$L_d \simeq L_1 \simeq L_2$ (and $\beta \simeq \beta_{cr}$) allows one to get the highest gain at the shortest
wavelength. At this point we have:

\begin{equation}
\lambar \simeq \epsilon \left( \frac{\gamma I_{\mathrm{A}}}{I} \right)^{1/4} \ ,
\label{lambar-opt-opt}
\end{equation}

\begin{equation}
L_d \simeq \epsilon \gamma_z^2 \left( \frac{\gamma I_{\mathrm{A}}}{I} \right)^{3/4} \ .
\label{drift-opt}
\end{equation}

\noindent The amplitude gain per cascade is given by (\ref{gain-l1}), the beta function  is given
by (\ref{beta-opt}), and the $R_{56}$ is given by (\ref{r56-opt}).

To estimate the total gain (and number of cascades)
required to reach saturation in LSCA, one has to estimate typical density
modulation for the start-up from shot noise. The power spectral gain of the amplifier depends on
the number of cascades $n$. For an optimal wavelength (\ref{lambar-opt}) as a central wavelength (
neglecting the dependence of the impedance on $k$),
and for the optimized $R_{56}$ from (\ref{r56-opt}), one easily obtains from (\ref{gain-wake}) that
the total power gain is proportional to $\hat{k}^{2n} \exp (-n \hat{k}^2)$, where $\hat{k}=k/k_{opt}$.
Thus, the relative bandwidth of the amplifier
is in the range 50-100 \%, depending on the number of cascades. Then an effective shot noise
density modulation can be estimated as \cite{we-micr}:

\begin{equation}
\rho_{sh} \simeq \frac{1}{\sqrt{ N_{\lambda}}} \ ,
\label{shot-rho}
\end{equation}

\noindent where $N_{\lambda} = I \lambda/(ec)$ is a number of electrons per central wavelength of
the amplified spectrum, $\lambda = 2 \pi \lambar$. At saturation the density modulation $\rho_{sat}$
is on the order of unity, so that the total amplitude gain $\rho_{sat}/\rho_{sh}$ is:

\begin{equation}
G_{tot} = G_1 G_2 ... G_n \simeq \sqrt{N_{\lambda}} \ ,
\label{gain-tot}
\end{equation}

\noindent where $G_n$ is the gain in the $n$-th cascade.
The total power gain of the saturated amplifier (that shows
an enhancement of power in a radiator with respect to spontaneous emission) is simply:

\begin{equation}
G_{tot}^{(p)} = G_{tot}^2 \simeq  N_{\lambda} \ .
\label{gain-tot-pow}
\end{equation}

We have presented a simple scheme for optimization of LSCA, but we should note that it is not strict and
serves for orientation in the parameter space only.
For instance, one can choose a drift length that is significantly shorter than the limit given by (\ref{ld-max})
and increase the number of cascades instead. In this case the formula (\ref{gain-ld}) should be used to
calculate the gain. For instance, three cascades with the gain 10 in each give about
the same total gain as two cascades with the gain 30 in each, but the total length of the amplifier
(for the same beta-function)
can be almost twice shorter\footnote{If one assumes that gain is the same in
all
cascades, gain per cascade is proportional to its length, and the length of a drift is always much
larger than the length of a chicane, then from
Eq.~(\ref{gain-tot}) one gets formally that the shortest total length is achieved at the number of
cascades $n \simeq \ln G_{tot}$, and the gain per cascade $G_n \simeq e \simeq 2.718...$. However, in
practice it is advisable to keep gain per cascade at least in the range $5-10$.}. If in addition one
reduces beta-function (since the drift space got shorter), one can go to shorter wavelength and higher
gain per cascade. So, if the gain per cascade at $\beta \simeq \beta_{cr}$ is very large, it could
be beneficial to go to the limit $L_d \simeq L_2 < L_1$ - if the beta-function is not getting too low
for technical realization.
On the other hand, in many cases even beta-function given by (\ref{beta-opt}) is too small
and technically not
feasible. In that case one would have to use larger values of $\beta$ and go to the limit defined by
plasma oscillations only, thus using Eqs.~(\ref{lambar-opt})-(\ref{gain-l1}).
If the wavelength of
interest differs significantly from the one given by (\ref{lambar-opt}), one should use
more general formulas of the previous Section.
In any case, the formulas of this Section are only estimates, and for
more accurate gain calculations one should use more general formulas of the previous Section (but then one
has to specify distributions), and in addition to include dynamics in drifts and chicanes more accurately.

\subsection{Wavelength compression}

As one can see from the formulas of the optimized LSCA,
a typical operating wavelength of an optimized LSCA is significantly longer than a wavelength that can be
reached in
SASE FELs (they can lase at $\lambar \simeq \epsilon$). In order to go to shorter wavelengths for given
electron beam parameters in LSCA, one would have to use wavelength compression. The broadband nature of
the amplifier makes this option especially attractive. Indeed, the compression factor is given by the formula:

\begin{equation}
C = (1-hR_{56})^{-1} \ ,
\label{comp-fac}
\end{equation}

\noindent where $h$ is the linear energy chirp (the derivative of relative energy slope). For a large $C$ a
variation of the compression factor reads:

\begin{equation}
\frac{\Delta C}{C} \simeq C \ \frac{\Delta h}{h} \ .
\label{comp-del}
\end{equation}

\noindent After the compression the bands of density modulations and of the radiator must overlap. This leads to
the following requirement on the compression stability:

\begin{equation}
\frac{\Delta C}{C} < \frac{\Delta k_{max}}{k} \ ,
\label{comp-del-1}
\end{equation}

\noindent where $\Delta k_{max}= max(\Delta k_{den},\Delta k_{rad})$, and $\Delta k_{den}$ and $\Delta k_{rad}$
are bandwidths of the density modulation and of the radiator, respectively. Thus, the stability of the chirp
must satisfy the requirement:

\begin{equation}
\frac{\Delta h}{h} < \frac{1}{C} \ \frac{\Delta k_{max}}{k} \ .
\label{h-del}
\end{equation}

For coherent FEL-type modulations and an undulator as a radiator $\Delta k_{max}/k \ll 1$ what might set very tight
tolerance for the chirp stability and limit practically achievable compression factors. For an LSCA, however,
$\Delta k_{max}/k = \Delta k_{den}/k \simeq 1$, so that for a given chirp stability one can go for much larger compression.
Alternatively, for a given compression factor one can significantly loosen the tolerances. Note also that
nonlinearities of the longitudinal phase space do not play a significant role in the case of LSCA.

If the
wavelength compression is applied, one should use formula (\ref{gain-wake}) to calculate gain per cascade,
and adjust $R_{56}$ to optimize the gain depending on compression factor $C$. One can consider different
options for compression. One possibility is to create an energy chirp before beam enters LSCA.
In this case one can adjust $R_{56}$ in different cascades in order to have a  mild compression in each cascade -
but this might shift the wavelength beyond the optimal range in the drifts of last cascades.
In that case one should make sure that the number of cascades and their parameters are adjusted
such that a saturation of LSCA is finally achieved.
Another possibility
is to create an energy chirp before the last cascade (for instance, by a laser in a short undulator
\cite{huang-24as}), and get
the desired compression in the last chicane. In this case, perhaps, one would need very strong energy chirp.
An interesting option for a short electron bunche with a high current would be to use an energy chirp,
induced by LSC along the whole bunch (then for compression one should use, for instance, doglegs instead of
chicanes, taking into account the sign of the energy chirp). In this case one should carefully adjust parameters
of the amplifier cascades since LSC (and a chirp) might strongly increase from one cascade to the next one.
However, loose tolerances (\ref{h-del}) can make such an option feasible.

\subsection{Undulator}

At the entrance of the undulator we have chaotically modulated electron beam with a typical amplitude of the order
of unity at saturation. The temporal correlations have the scale of a wavelength, and the spectrum is broad.
The undulator radiation
within the central cone\footnote{There is also emission of powerful radiation beyond the central cone.}
$\sqrt{\lambda/L_{w}}$ (here $L_{w}$ is the undulator length) has a relative bandwidth
$N_{w}^{-1} \ll 1$, where $N_{w}$ is the number of undulator periods. In the case when Fresnel number is small, $\sigma_{\bot}^2/(\lambar L_{w}) \ll 1$, the radiation power
within the central cone is equal to the power of spontaneous emission multiplied by the power gain $N_{\lambda}$ of the LSCA.
In this limit the power does not depend on the number of undulator periods.
One can easily see that the Fresnel number is always small if the condition (\ref{lambar-opt}) is satisfied,
transverse size of the beam in the undulator is the same as that in amplification cascades, and
there is no wavelength compression. In this case the transverse coherence is guaranteed.
However, with a strong wavelength compression in the last chicane of LSCA, the Fresnel number might no
longer be small, so that one
should use more general formula for the radiated power \cite{nima-replica}. In that case the transverse
coherence can still be relatively good if in the drifts of LSCA the wavelength is given by the condition
(\ref{lambar-opt}) or it is longer. Indeed, transverse correlations are established in this case
due to LSC \cite{venturini}. We do not discuss in this paper nonlinear harmonic generation in LSCA, since
it would be highly speculative without numerical simulations. We can only mention here that this
should be possible in a saturated LSCA.

\subsection{Numerical example}

Let us illustrate the formulas of this Section with a numerical example. We consider the electron beam
with the following parameters: the energy 3 GeV ($\gamma \simeq 6 \times 10^3$), peak current 2 kA,
normalized emittance 2 $\mu$m, rms energy spread 0.3 MeV ($\sigma_{\gamma} = 0.6$). The
energy modulations due to LSC are accumulated in the focusing channels,
so that $\gamma_z = \gamma$ (no undulators are placed there). From the formula (\ref{beta-opt})
we find $\beta_{cr} \simeq 1.4$ m. Although
this low value assumes that the focusing channels are densely packed with the quadrupoles, it is technically
possible. From (\ref{lambar-opt-opt}) or (\ref{lambar-opt}) we get $\lambar \simeq 2.5$ nm, i.e.
the wavelength $\lambda$ is about 15 nm. The optimal $R_{56}$ for this wavelength is about 25 $\mu$m
(and the chicane may fit in the space between two quadrupoles, so that periodicity of the channel
is not disturbed) .
According to (\ref{drift-opt}), the length of the drift space is about 20 m.
The gain per cascade can be found from (\ref{gain-l1}), it is larger than 40. According to (\ref{gain-tot}),
the total gain should be about $10^3$, so that only two cascades are sufficient. As it was mentioned before,
such a high gain per cascade is not optimal from the point of view of the total length of the system, and 3-4
cascades would be more preferable. For instance, for 3 cascades one needs the gain per cascade about 10,
so that the length of a cascade would be about 5 m. The chicanes can be made very compact. Behind the last
chicane a tunable-gap undulator with the period length of 5 cm and a number of periods 30 can be installed. The
total length of this system is less than 20 m. The undulator selects a relatively narrow band of
about 3 \% from the broad-band density modulations. The peak power within the central cone is estimated
at a gigawatt level, assuming that amplification of density modulation reached saturation in the last
cascade. As it was discussed before, a shift of the central wavelength by a factor of 2 does not change the impedance
significantly. The gain is also affected weakly as soon as the $R_{56}$ is tuned correspondingly. The
undulator wavelength is adjusted by tuning the gap. Therefore, wavelength tunability in the range
7-30 nm is easily possible without increasing the total length of the system.

This numerical example illustrates that LSCA can not directly compete with FELs in
terms of shorter wavelengths (although the wavelength compression can help),
higher power and brilliance etc. For instance, a SASE FEL with the given
electron beam parameters could successfully saturate at 1-2 nm wavelength within 30-50 m of undulator
length and produce a few gigawatts of peak power within a bandwidth that is smaller than a per cent.
However, LSCA can be a cheap solution for generation
of longer wavelength radiation with a relatively high beam energy,
and can have other attractive applications that are discussed in the following Sections.

\section{Possible applications of LSCA}

\subsection{LSCA as a cheap addition to existing or planned X-ray FELs}

Undulator beamlines of the existing and planned X-ray FELs often consist of long drift spaces and long undulators.
Insertion of a few chicanes and a short undulator at the end may allow for a parasitic production of relatively
long wavelength radiation (as compared with the FEL wavelength) by the same electron bunch. This would extend in an
inexpensive way the wavelength range of a facility. Moreover, since both radiation pulses are perfectly synchronized,
they can be used in pump-probe experiments.

As a first example let us consider the undulator beamline SASE1 at the European XFEL \cite{euro-xfel-tdr}. There is a long
drift space (about 300 m) in front of SASE1 undulator, and 200 m long drift behind the undulator. The undulator itself
has the total length of 200 m (magnetic length 165 m plus 35 meters of intersections). Let us consider the electron beam with the following parameters:
energy 17.5 GeV, normalized slice emittance 0.4 mm mrad, peak current 3-4 kA, slice energy spread 1.5 MeV.
The tunable-gap undulator
is assumed to be tuned to the resonance with the wavelength 0.05 nm, so that $\gamma_z = 1.9 \times 10^4$.
The optimal beta-function in the undulator for
these beam parameters is about 15 m, and it is about 30-40 m in the drifts. The core of the bunch with high current saturates
at the FEL wavelength in the undulator, so that this part of the bunch is spoiled (has a large energy spread).
We consider parts of the bunch with the current about 1 kA assuming that there is no FEL saturation there.
We propose to install three compact chicanes just in front of the undulator, just behind it, and at the end of the
second drift. Thus, we have three amplification cascades of LSCA that operates parasitically.
The last chicane is followed by a short undulator.
From the formulas of the previous Section
we find that the optimal wavelength for LSC instability is $\lambda \simeq 4$ nm. The optimal $R_{56}$ is about
8 $\mu$m for all cacsades.
Beta-function in all cascades is much larger than $\beta_{cr}$, moreover the lengths of all cascades are shorter
than reduced wavelength of plasma oscillations, i.e. $L_d < L_1$. Therefore, we use formula (\ref{gain-ld}) to calculate
gain in every cascade. We find that the total gain is given by the following product of partial gains:
$G_{tot} \simeq 8 \times 13 \times 5 \simeq 500$. This is larger than the gain required to reach saturation, about 300
according to (\ref{gain-tot}). We choose an undulator with 50 periods and a period length 10 cm. Radiation power within
the central cone exceeds that of spontaneous emission by 5 orders of magnitude and is in sub-GW level with the
bandwidth about 2 \%, radiation is transversely coherent.
The tunability can be easily
achieved in the range of 2-10 nm by changing the $R_{56}$ and the undulator gap. The soft X-ray pulses are synchronized
with hard X-ray pulses produced by
the core of the same bunch, so that these two colors can be used in pump-probe experiments. Alternatively, they can
be separated and used by different experiments\footnote{As an option one can consider the bending system (with properly
adjusted $R_{56}$) between SASE1 and
the downstream soft X-ray undulator SASE3 \cite{euro-xfel-tdr} as an alternative to the last chicane.
Then the short undulator is placed in SASE3 beamline thus extending its wavelength range.}.

Note that in this example we considered a parasitic use of the beamline and of an unspoiled part of the
electron bunch. With a dedicated use of the high-current part of the bunch one can essentially reduce the
total length of the amplifier. Let us consider the same electron bunch as before, but now we assume that
the core of the bunch with the current 3 kA is not spoiled by FEL interaction (for instance, some
bunches are kicked in front of the undulator by the fast kicker \cite{low-em}).
We consider an operation of LSCA in the drift behind the undulator,
requiring beta-function to be about 10 m (somewhat larger that $\beta_{cr}$),
thus the optimal wavelength is 2 nm.
Choosing length of the drift in an amplification cascade to be 30 m (it is much smaller than $\lambar_p$),
and the $R_{56} \simeq 4 \ \mu$m, we find with the help of (\ref{gain-ld}) that the gain per cascade is
about 5. To reach saturation one would need four cascades, so that the total length of the amplifier would
be about 120 m. A gigawatt-level radiation power would then be produced within the central cone of a short
undulator, tunability between 1 nm and 5 nm is easy to obtain.

Parasitic use of long drifts and unspoiled parts of an electron beam is possible at other facilities,
for example, at the soft X-ray FEL user facility FLASH \cite{flash-nat-phot,njp}.
There is about 45 m long
drift space in front of the 27 m long undulator. Without going into the details, we notice that by installing
two chicanes (in front of the undulator and behind it, $R_{56} \simeq 200 \ \mu$m) and a short radiator undulator
one can parasitically generate powerful VUV radiation with the wavelength around 100 nm.

\subsection{Generation of attosecond pulses}

There are many proposals to produce attosecond pulses from FELs
\cite{attofel-oc,oc-2004-2,atto-b,atto-e,atto-f,prstab-2006-2,huang-24as}.
In principle, by using strongly nonlinear
manipulations with the longitudinal phase space, one can reduce X-ray pulse duration down to several
cycles \cite{huang-24as}. Here we note that the broadband nature of the LSC instability suggests that few-cycle
pulces can be naturally produced in LSCA. There might be different solutions, we consider the one similar to
the current-enhanced SASE scheme \cite{atto-f}. The main idea is that a very short slice (on the order of 100 as)
is created in the electron bunch such
that the current in this slice is much higher than that in the rest of the bunch. The FEL saturation in this slice
is achieved earlier so that the power in it is much higher. This local enhancement of the current is achieved due
to the modulation of the electron beam in energy by few-cycle laser in a short undulator, and then by using
a chicane.

Let us consider again SASE1 beamline at the European XFEL \cite{euro-xfel-tdr}. Now we assume
a dedicated mode of operation
of the LSCA (and no SASE operation) with the goal of production of the attosecond pulses. A relatively low-current beam
(peak current about 100 A)\footnote{Beam can be compressed in a single bunch compressor or with the help of
velocity bunching.} is required for the operation of this scheme.
Somewhere at the beginning of the first drift we modulate the beam by 5 fs long pulse from
Ti:S laser system in the two-period undulator. Amplitude of the energy modulation is about 3 MeV. In the chicane
with the $R_{56} \simeq 0.6$ mm we obtain a spike with the current about 1 kA and the rms width about 10 nm.
The amplification of shot noise within this spike to saturation takes place in three amplification cascades as
described above, in the example with the parasitic use of SASE1 beamline.
The soft X-ray pulses (in the range 2-5 nm) with the duration about 100 as and peak power
a few hundred megawatts are produced in the undulator with the number of periods from five to ten.
Note that LSC not only plays a positive role in amplification mechanism, it also induces an energy chirp
along the spike. In this scheme it cancels first the linear part of a chirp induced by a laser, and then
it induces the chirp of the opposite sign but of the same order. These chirps
lead to a weak compression (decompression) in the chicanes but do not disturb the operation of the scheme.

\subsection{LSCA as a source of radiation with a relatively large bandwidth}

FEL radiation has narrow band, typically 0.1-1 \%. For some experiments, however, a relatively large bandwidth
is required, up to 10 \%. In FELs an increase can be achieved by imposing on the electron beam an energy chirp,
which is translated into radiation frequency chirp.
This approach has technical limits: accelerator has a finite energy acceptance, and it is not always possible to
impose required energy chirps on very short bunches. In an LSCA the density modulation is broadband, and the radiation
bandwidth (within the central cone of undulator radiation) is given by inverse number of undulator periods.

\subsection{LSCA driven by a laser-plasma accelerator}

The technology of laser-plasma accelerators progresses well \cite{esarey},
a GeV beam is already obtained \cite{leemans}.
The electron beam with the energy about 200 MeV was sent through the undulator, and spontaneous undulator radiation
at 18 nm wavelength was obtained \cite{fuchs}. The VUV and X-ray FELs driven by these accelerators are proposed
\cite{gruener,fawley}.
However, it is not clear at the moment if tight requirements on electron beam
parameters and their stability, overall accuracy of the system performance etc., could be achieved
in the next years.

Contrary to FELs, the amplification mechanism of LSCA is very robust.
For example, it can tolerate large energy chirps. In the case of an FEL the
energy chirp parameter is $\lambar h/\rho^2$, where $\rho$ is the FEL parameter \cite{boni-rho} defining, in particular,
SASE FEL bandwidth. The energy chirp parameter should be small as compared to unity in order to not affect FEL gain.
Contrary to that, mechanism of LSCA is broadband, so that "effective $\rho$" is on the order of one. In other
words, in a drift space the influence of the chirp can be always neglected. Of course,
if one would like to avoid compression (decompression) in chicanes of LSCA, one should require $h R_{56} \ll 1$.
If the $R_{56}$ is chosen according to (\ref{r56-opt}), then the condition for the chirp can be formulated
as $h \lambar \ll \sigma_{\gamma}/\gamma$.

In an FEL there are stringent requirements on straightness of the trajectory: the electron beam
must overlap with radiation over a long distance. In the case of LSCA one should only require that
the angles of the electron orbit should be smaller that $\lambar/\sigma_{\bot}$ what means for the optimal wavelength
$\gamma_z^{-1}$.

One can speculate (since some important parameters of beams have never been measured)
that LSCA could be an interesting alternative to FELs, at least as the first step towards
building light sources based on laser-plasma accelerators. One of the most important unknown parameters is
the slice energy spread (slice size is given by a typical wavelength amplified in LSCA),
since the measured value is
usually a projected energy spread, dominated by an energy chirp along the bunch.
An interesting option would be to use an energy
chirp, induced by LSC and wakefields over the whole bunch \cite{geloni-plasma,gruener},
for the wavelength compression as discussed
in Section 2. Taking into account the sign of the energy chirp, one should use, for instance doglegs
instead of chicanes. One can also consider LSCA as a preamplifier (making sure that it does not saturate)
with the final amplification in an FEL.

\section{Discussion}

In this paper we introduced the concept of the Longitudinal Space Charge Amplifier (LSCA) that can operate
in VUV and X-ray ranges. Although such an amplifier
can not directly compete with FELs in terms of shorter wavelength, higher power, brilliance etc., one can
nevertheless find interesting applications for it. In particular, it can be a cheap addition to some existing or
planned FEL systems helping to extend operating range towards longer wavelength and to provide the second color
for pump-probe experiments. Broadband nature of the amplifier supports production of short (down to few cycles)
VUV and X-ray pulses. Bandwidth of the radiation from the undulator of LSCA can be controlled by
choosing the number of undulator periods. In particular, one can produce powerful radiation with a relatively large
bandwidth what might be difficult in an FEL. Robustness of LSCA makes it an interesting alternative to an FEL in
light sources driven by laser-plasma accelerators.

There are many different possibilities that were not considered in this paper and are left for future
studies. In particular, we did not study nonlinear harmonic generation, an effect that should occur
at the saturation of LSCA. Since amplification mechanism of LSCA is broadband, the bands of harmonics
might even partially overlap.
The radiation wavelength within the central cone is controlled by tuning the undulator parameter.
Also, for a planar undulator there might be a set of odd harmonics on axis.

We have considered in this paper a start up of LSCA from shot noise.  However, LSCA can also amplify a
coherently seeded density modulation. In this case one needs an undulator and a chicane in front of the first
cascade of LSCA. The electron beam gets energy-modulated by a laser beam in the undulator, and in the chicane
these energy modulations are converted into coherent density modulations. Particularly interesting might be
a seeding in a few-period undulator by attosecond pulses, obtained by high harmonics generation (HHG)
in gases by powerful few-cycle lasers \cite{krausz}. Short few-cycle density modulations can be
amplified through
LSCA without lengthening, and few-period radiator undulator would produce powerful few-cycle VUV radiation.
This option is not available in an FEL amplifier due to a narrow bandwidth.

It was briefly mentioned in the paper that LSCA can serve as a preamplifier for a SASE FEL.
There might be other options, for instance putting LSCA with a short undulator in an optical cavity,
thus having, for instance, a regenerative amplifier with a desirable bandwidth.
We should also notice here a possibility of using LSCA for some amplification (not necessarily to
saturation) of shot noise in light sources based on spontaneous radiation in undulators, for example
driven by energy-recovery linacs. In this way one can significantly enhance radiation intensity and brilliance.

Finally, we have to mention that a possibility of a harmful LSC instability at short wavelengths
(VUV and soft X-ray)
should not be forgotten. Such an instability can develop parasitically in FEL systems (at wavelengths that are
much longer than the FEL wavelength) with dispersive elements,
such as chicanes in high-gain harmonic generation schemes (especially dangerous can be "fresh bunch" chicanes),
achromatic bends for separation of beamlines, chicanes in seeding and self-seeding schemes etc.
If LSC instability develops to a significant level of density modulations, strong energy modulations
(acting as local energy spread) can be induced in last parts of FELs thus hampering their operation.

\section{Acknowledgements}

We would like to thank R.~Brinkmann, P.~Emma, and Z.~Huang for useful discussions.

\end{document}